\documentclass[12pt]{article}
\usepackage{graphicx}% Include figure files
\usepackage{amssymb}
\usepackage{amsmath}
\usepackage{color}
\newcommand{\eV}{\mbox{~eV}}
\newcommand{\keV}{\mbox{~keV}}

\newcommand{\GeV}{\mbox{~GeV}}

\newcommand{\eqn}[1]{&\hspace{-0.6em}#1\hspace{-0.6em}&}
\newcommand{\FN}{\rm FN}
\newcommand{\BL}{\rm B-L}
\begin{document}
\baselineskip 0.6cm
%
%%%%%%%%%%%%%%%%%%%%%%%%%%%%%%%%%%%%%%%%%%%%%%%%%%%%%%%%%%%%%%%%%%%%
%%%%%  Title Page  %%%%%%%%%%%%%%%%%%%%%%%%%%%%%%%%%%%%%%%%%%%%%%%%%
\begin{titlepage}
\begin{center}

%%%%%%%% Preprint #
\begin{flushright}
SU-HET-09-2014
\end{flushright}

\vskip 2cm

%%%%%%%% Title
{\Large \bf 
{\boldmath $\nu_R$} dark matter-philic Higgs 
for 3.5 keV X-ray signal
}

\vskip 1.2cm

%%%%%%%% Authors
{\large 
Naoyuki Haba, Hiroyuki Ishida and Ryo Takahashi
}

\vskip 0.4cm

%%%%%%%% Addresses
{\em
  Graduate School of Science and Engineering, Shimane University, Matsue, 690-8504 Japan
}

\vskip 0.2cm

%%%%%%%% Date
%(\today)
%(April 30, 2010)

\vskip 2cm

\vskip .5in
%%%%%%%%%%%%%%%% Abstract %%%%%%%%%%%%%%%%%%%%%%%%%%%%%%%%%%%%%%%%%%
\begin{abstract}
We suggest a new model of $7$ keV right-handed neutrino dark matter 
inspired by a recent observation 
of $3.5 {\keV}$ X-ray line signal in the XMM-Newton observatory.
It is difficult to derive the tiny masses with a suitable left-right mixing of the neutrino 
in a framework of ordinary simple type-I seesaw mechanism.
We introduce a new Higgs boson, a dark matter-philic Higgs boson, 
in which the smallness of its vacuum expectation value can be achieved.
We investigate suitable parameter regions 
where the observed dark matter properties are satisfied.
We find that the vacuum expectation value of 
dark matter-philic Higgs boson should be about $0.17\GeV$.
\end{abstract}
%%%%%%%%%%%%%%%% Abstract %%%%%%%%%%%%%%%%%%%%%%%%%%%%%%%%%%%%%%%%%%
%%%%%%%%%%%%%%%%%%%%%%%%%%%%%%%%%%%%%%%%%%%%%%%%%%%%%%%%%%%%%%%%%%%%
\end{center}
\end{titlepage}
%%%%%  Title Page  %%%%%%%%%%%%%%%%%%%%%%%%%%%%%%%%%%%%%%%%%%%%%%%%%

%%%%%%%%%%%%%%%%%%%%%%%%%%%%%%%%%%%%%%%%%%%%%%%%%%%%%%%%%%%%%%%%%%%%
\renewcommand{\thefootnote}{\#\arabic{footnote}} 
\setcounter{footnote}{0}
%\clearpage
%%%%%%%%%%%%%%%%%%%%%%%%%%%%%%%%%%%%%%%%%%%%%%%%%%%%%%%%%%%%%%%%%%%%
%%%%% ** Text ** %%%%%%%%%%%%%%%%%%%%%%%%%%%%%%%%%%%%%%%%%%%%%%%%%%%
%%%%%%%%%%%%%%%%%%%%%%%%%%%%%%%%%%%%%%%%%%%%%%%%%%%%%%%%%%%%%%%%%%%%
%
\section{Introduction}
The existence of dark matter (DM) is 
confirmed by a lot of astrophysical observations. However, there is no 
candidate in the standard model (SM) particle content.  
One of properties required for DM is a long lifetime 
which should 
be longer than the age of our universe.
To realize this stability, 
an additional symmetry should be exist. 
For example, when the DM particle has odd parity and others do even parity 
under a $Z_2$ symmetry, the DM particle can be stable. 

Recent analyses of the observation of X-ray from the Perseus galaxy clusters 
and the Andromeda galaxy by the XMM-Newton observatory reported an unknown 
X-ray line spectrum around $3.5 {\keV}$ \cite{Bulbul:2014sua,Boyarsky:2014jta}. 
The origin of this line cannot be explained by astrophysical phenomena so far.
Therefore, we try to explain the X-ray line spectrum in the particle physics. 
An interesting explanation is given by decay of a light DM particle.
After these announcements, a lot of articles \cite{Ishida:2014dlp}-\cite{Dutta:2014saa} 
in which DM decays into a photon 
have been researched.
One attractive candidate for a decaying DM is the right-handed neutrino, $\nu_{R}$, 
with mass around $7 \keV$\footnote{See \cite{Merle:2013gea} for a review of keV DM 
model building.}. In this scenario, the experiments show 
the mixing angle between left- and right-handed neutrinos, 
$\Theta$, to be 
$\Theta^2=(0.55~\mathchar`-~5.0) \times 10^{-11}$.

On the other hand, it is well known that the right-handed neutrinos 
can generate the tiny neutrino masses through the type-I seesaw 
mechanism~\cite{seesaw}. 
The Lagrangian for the seesaw mechanism is described as
\begin{eqnarray}
\mathcal{L}_{\rm seesaw} \eqn{=} 
- y_{\alpha I} \bar{L}_\alpha \tilde{H} \nu_{RI} 
+ \frac{M_N}{2} \bar{\nu}_{RI}^c \nu_{RI}\,,\label{Eq:seesaw}
\end{eqnarray}
with 
$\tilde{H}=i \sigma_2 H^\ast$. 
$L$ and $H$ are lepton and Higgs doublets, respectively. 
Here, $M_N = {\rm diag} (M_1\,,M_2\,,M_3)$ is the Majorana mass matrix of the 
right-handed neutrinos. 
$\alpha = e\,,\mu\,,\tau$ and $I=1\, \mathchar`-\, 3$ represent the flavor of 
left-handed leptons and right-handed neutrinos, respectively. 
If the right-handed neutrinos have a large mass hierarchy, 
it can explain several issues such as the tiny neutrino masses, DM, and baryon 
asymmetry of the universe simultaneously ({\it e.g.} 
see~\cite{Asaka:2005an,Kusenko:2010ik,Barry:2011fp,Chen:2011ai}). Such a 
hierarchical structure among the Majorana masses can be realized by {\it e.g.,} the 
Froggatt-Nielsen mechanism \cite{Froggatt:1978nt,Merle:2011yv}, the split seesaw
 mechanism \cite{Kusenko:2010ik}, a lepton flavor symmetry~\cite{Lindner:2010wr}, 
an extra-dimensional extension~\cite{Takahashi:2013eva} and an extended B-L 
structure~\cite{Ishida:2013mva}, and so on. 
%We show why a simple seesaw mechanism does not work below.

However, once the right-handed neutrino with the mass of order keV 
gives the solar or atmospheric neutrino mass scale through the type-I seesaw mechanism, 
the size of left-right mixing angle cannot be explained.
After breaking the electroweak symmetry, 
we have the Dirac mass term,
 $(M_D)_{\alpha I}=y_{\alpha I} v$ with $v \equiv \langle H \rangle$, and the 
Majorana mass term. 
The seesaw mechanism can explain the tiny neutrino masses as $m_\nu = -M_D^2/M_N$ 
with $m_\nu \sim \mathcal{O} (0.1) \eV$ even 
under hierarchical mass spectra of Majorana masses. 
The mixing angle between the left- and right-handed neutrinos is given by  
$\Theta_{\alpha I}=(M_D)_{\alpha I}/M_I$. 
When the lightest Majorana mass is of order ${\rm keV}$, 
the seesaw relation induces the left-right mixing angle as 
\begin{eqnarray}
\Theta^2
\eqn{=} 
\sum_\alpha \Theta_{\alpha 1}^2 
\simeq
\left( \frac{M_D}{M_1} \right)^2 
= 
\frac{m_\nu}{M_1} \sim 10^{-4} \gg 10^{-11}\,.\label{Eq:FN}
\end{eqnarray}
It is worth noting that the suitable magnitude of the mixing angle, $\Theta^2 \sim 10^{-11}$, never achieves unless giving up the seesaw relation.
It is because a tiny neutrino Yukawa couplings only for the first generation 
with the left-right mixing angle, $\Theta^2 \sim 10^{-11}$ induces 
the mass scale of $(y_{\alpha 1}v)^2/M_1$, which is much smaller than the atmospheric  neutrino mass scale, $m_\nu \sim \mathcal{O} (0.1) \eV$.

In this work, we suggest a new possibility to realize the suitable left-right mixing due to  
the deviation from the seesaw relation by an extending SM. 
We introduce a DM-philic Higgs field, 
$H_{\rm DM}$, which constitutes a Yukawa interaction of DM with ordinary SM 
fields.

\section{DM-philic Higgs model}
\subsection{Basic idea}

The results from X-ray observation experiments indicate 
the existence of an unknown particle which has a mass around $7 \keV$.
If it is a right-handed neutrino, 
it should not contribute to the active neutrino mass scale through the seesaw relation 
due to its tiny mass and the suitable mixing angle 
as discussed above.
Such property could imply 
the right-handed neutrino DM (RH$\nu$DM) has a different interaction 
from other particles. 
We introduce an extra field: a RH$\nu$DM-philic Higgs boson, $H_{\rm DM}$.

By introducing $H_{\rm DM}$, the relevant Lagrangian is given by 
\begin{eqnarray}
\mathcal{L}_{\rm seesaw}' 
\eqn{=} 
- y_{\alpha 1} \bar{L}_\alpha \tilde{H}_{\rm DM} \nu_{R1} 
- y_{\alpha i} \bar{L}_\alpha \tilde{H} \nu_{Ri} 
+ \frac{M_N}{2} \bar{\nu}_{RI}^c \nu_{RI}\,,\label{Eq:simple}
\end{eqnarray}
where $i=2\,,3$ and we suppose that $M_1 \ll M_{i}$ at this moment.
Therefore, we can realize a small mixing of $\nu_{R1}$ 
by taking small vacuum expectation value (VEV), $v_{\rm DM}$.
Under this simple setup, 
the mixing angle can be represented as
\begin{eqnarray}
\Theta^2 \eqn{\simeq} 
\left( \frac{y_{\alpha 1} v_{\rm DM}}{M_1} \right)^2 
=
\left( \frac{v_{\rm DM}}{v} \right)^2 \left( \frac{y_{\alpha 1} v}{M_1} \right)^2 
= 
\left( \frac{v_{\rm DM}}{v} \right)^2 \frac{m_\nu}{M_1} 
\sim \left( \frac{v_{\rm DM}}{v} \right)^2 \times 10^{-4}
\,,\label{Eq:dev}
\end{eqnarray}
where the third and fourth equalities are given by Eq. (\ref{Eq:FN}).
One can see that when $\left( v_{\rm DM}/v \right)^2$ is $\mathcal{O} (10^{-7})$, 
the observed X-ray flux can be explained.
Therefore, $v_{\rm DM}$ is needed to be $\mathcal{O} (0.1) \GeV$.
Instinctively, the deviation from the seesaw formula is realized 
by the small value of $v_{\rm DM}$ rather than a fine-tuning of coupling constants.
This is our main idea and this possibility is really simple.

\subsection{A simple model}
Here, we show an example of realistic model.
The charge assignment of each field is summarized in Tab.~{\ref{Tab:assign}}.
\begin{table}[t]
\begin{center}
\renewcommand{\arraystretch}{1.5}
\begin{tabular}{|c||c|c|c|c|c|c|}
\hline
 &$H$ &$\tilde{H}_{\rm DM}$ &$\Phi_{\FN}$ &$\Phi_{\BL}$ &$\nu_{R1}$ &$\nu_{Ri}$\\
\hline
$U(1)_{\BL}$ &0 &0 &0 &$2$ &$-1$ &$-1$\\
\hline
$U(1)_{\FN}$ &0 &0 &$-1$ &0 &$n$ &0\\
\hline
$Z_{2}$ &+ &$-$ &+ &+ &$-$ &+\\
\hline
\end{tabular}
\caption{The charge assignments for each field. $n$ is an arbitrary natural number.}\label{Tab:assign}
\end{center}
\end{table}
We also introduce gauged $U(1)_{\BL}$ 
which is responsible not only for the generation of the Majorana masses of RH$\nu$s
but also the DM production.
Two global symmetries $U(1)_{\FN}$ and $Z_2$ 
provide the sufficient mass hierarchy and couplings between $\nu_{R1}$ and others.
$U(1)_{\BL}$ and $U(1)_{\FN}$ are spontaneously broken by a B$-$L Higgs boson, $\Phi_{\BL}$ and another Higgs boson, $\Phi_{\FN}$, respectively.
Those breaking scales are denoted as $v_{\BL}$ and $v_{\FN}$, respectively.

The Lagrangian under these symmetries are given as
\begin{eqnarray}
\mathcal{L} \eqn{=} \mathcal{L}_{\rm SM} + \mathcal{L}_{\nu}\,,\\
\mathcal{L}_\nu \eqn{=} 
y_{\alpha i} \bar{L}_\alpha \tilde{H} \nu_{Ri} 
+ \lambda_{ij} \Phi_{\BL} \bar{\nu}_{Ri}^c \nu_{Rj} 
+ \frac{\lambda_{\alpha 1}}{\Lambda^n} \Phi_{\FN}^n \bar{L}_\alpha \tilde{H}_{\rm DM} \nu_{R1} 
+ \frac{\lambda_{11}}{\Lambda^{2n}} \Phi_{\FN}^{2n} \Phi_{\BL} \bar{\nu}_{R1}^c \nu_{R1}\,,
\end{eqnarray}
where $\mathcal{L}_{\rm SM}$ contains the terms under the SM gauge symmetries, 
$y$ and $\lambda$ are dimension-less coupling constants, and 
$\Lambda$ represents a cutoff scale.

Next, we consider the Higgs potential. 
For the realistic model, the small magnitude of $v_{\rm DM}$ is essential.
We show the validity of this point below.
The corresponding Higgs potential of this model is 
 \begin{eqnarray}
  V &=& m_1^2|H|^2+m_2^2|H_{\rm DM}|^2-m_3^2(H^\dagger H_{\rm DM}+H_{\rm DM}^\dagger H)
        +\frac{\lambda_1}{2}|H|^4+\frac{\lambda_2}{2}|H_{\rm DM}|^4 
        \nonumber \\
    & & 
        +\lambda_3|H|^2|H_{\rm DM}|^2
        +\lambda_4|H^\dagger H_{\rm DM}|^2 
        + \frac{\lambda_5}{2} \left[ (H^\dagger H_{\rm DM})^2 + (H_{\rm DM}^\dagger H)^2 \right]
\,,\label{v}
\end{eqnarray}
where $m_3^2(H^\dagger H_{\rm DM}+H_{\rm DM}^\dagger H)$ terms break 
$Z_2$ symmetry softly.
Thus, the lightest RH$\nu$ becomes DM which can decay. 
A stationary condition $\partial V/\partial v_{\rm DM}=0$ gives 
 \begin{eqnarray}
  v_{\rm DM}\simeq
  \frac{m_3^2}{m_2^2} v, \label{Eq:vDM}
 \end{eqnarray}
where we assume 
$
\sqrt{\lambda_3}v,\sqrt{\lambda_4}v\ll m_2
$. 
Note that 
a tiny value of $v_{\rm DM}$ can be realized 
when $m_3\ll m_2$ in Eq. (\ref{Eq:vDM}).

After each scalar field takes its VEV, the Dirac and Majorana mass matrices can be written as
\begin{eqnarray}
M_D \eqn{=}
\begin{pmatrix}
\frac{\lambda_{11}}{\Lambda^n} v_{\FN}^n v_{\rm DM} &y_{12} v &y_{13} v\\
\frac{\lambda_{21}}{\Lambda^n} v_{\FN}^n v_{\rm DM} &y_{22} v &y_{23} v\\
\frac{\lambda_{31}}{\Lambda^n} v_{\FN}^n v_{\rm DM} &y_{32} v &y_{33} v
\end{pmatrix}\,,\label{Eq:Dirac}\\
M_M \eqn{=}
\begin{pmatrix}
\frac{\lambda_{11}}{\Lambda^{2n}} v_{\BL} v_{\FN}^{2n} &0 &0\\
0 &\lambda_{22} v_{\BL} &\lambda_{23} v_{\BL} \\
0 &\lambda_{32} v_{\BL} &\lambda_{33} v_{\BL} 
\end{pmatrix}\,.\label{Eq:Majorana}
\end{eqnarray}
The hierarchical structure of the Majorana masses as $M_1 \ll M_{i}$ 
and the different coupling property in the Dirac mass term are realized.

The seesaw mechanism provides the mass matrix for the active neutrinos, 
$M_\nu$, as 
\begin{eqnarray}
(M_\nu)_{\alpha \beta} \eqn{=} 
\lambda_{\alpha 1} \lambda_{\beta 1} \frac{v_{\FN}^{2n} v_{\rm DM}^2}{\Lambda^{2n} M_1} 
+ 
\sum_i y_{\alpha i} y_{\beta i} \frac{v^2}{M_i} \,,
\end{eqnarray}
where $M_i$ are obtained by the diagonalization of Eq. (\ref{Eq:Majorana}).
$M_\nu$ can be diagonalized by the so-called MNS matrix, $U$, 
as $M_\nu^{\rm diag} = {\rm diag} (m_1\,,m_2\,,m_3)= U^\dagger M_\nu U^\ast$.
Roughly speaking, the order of $m_I$ can be estimated 
in the normal mass ordering case as 
\begin{eqnarray}
m_1 \eqn{\simeq} 
\lambda_{\alpha 1} \lambda_{\beta 1} \frac{v_{\FN}^{2n} v_{\rm DM}^2}{\Lambda^{2n} M_1} 
= \frac{\lambda_{\alpha 1} \lambda_{\beta 1}}{\lambda_{11}} \frac{v_{\rm DM}^2}{v_{\BL}}\,,\\
m_{i} \eqn{\simeq} 
y_{\alpha i} y_{\beta i} \frac{v^2}{M_i} = \frac{y_{\alpha i} y_{\beta i}}{\lambda_{ij}} \frac{v^2}{v_{\BL}}\,,
\end{eqnarray}
where $m_3$ is of $\mathcal{O} (0.1) \eV$ 
which corresponds to the atmospheric neutrino mass scale and 
the squared difference between $m_2$ and $m_1$ explains the solar mass-squared difference.
Thus, the $U(1)_{\BL}$ breaking scale can be determined as $v_{\BL} \simeq 3.0 \times 10^{14}\GeV$. 
Note that we can reproduce the observed mixing angles in the MNS matrix 
by taking proper values of neutrino Yukawa couplings, $y_{\alpha \beta}$ 
\footnote{We can also reproduce the inverted mass ordering case.}.

\section{Phenomenological consequence}
We are at the position to investigate suitable model parameters by cosmological observations.
We assume that all of the dimension-less couping constants are unity just for simplicity 
and the cutoff scale is the reduced Planck scale, $\Lambda = 2.4 \times 10^{18} \GeV$. 

Firstly, we estimate the breaking scale of the extra symmetries 
by the lightest right-handed neutrino mass, $M_1$, 
which is given by Eq.(\ref{Eq:Majorana}) as 
\begin{eqnarray}
M_1 \simeq 
\left( \frac{v_{\FN}}{\Lambda} \right)^{2n} v_{\BL}\,.
\end{eqnarray}
The left-right mixing angle of the neutrinos is estimated as 
\begin{eqnarray}
\Theta^2 \eqn{\simeq} \left( \frac{(M_D)_{\alpha1}}{M_1} \right)^2 
\simeq 
\left( 
\frac{v_{\FN}^n v_{\rm DM}/\Lambda^n}{v_{\FN}^{2n} v_{\BL}/\Lambda^{2n}}
\right)^2\,.\label{Eq:mixing}
\end{eqnarray}
From the observations, each value is constrained as \cite{Boyarsky:2014jta}
\begin{eqnarray}
M_1 \eqn{=} 7.06 \pm 0.05 \keV\,,\\
\Theta^2 \eqn{=} (0.55~ \mathchar`-~ 5.0) \times 10^{-11}\,,
\end{eqnarray}
and here, we fix these as $M_1 = 7.06\keV$ and 
$\Theta^2 = 1.3 \times 10^{-11}$ which are the best fit values of the experiments.

We first mention to the scale of $v_{\rm DM}$.
One can rewrite Eq. (\ref{Eq:mixing}) by the seesaw neutrino mass, $m_\nu$, as
\begin{eqnarray}
\Theta^2 
\eqn{\simeq}
\left( 
\frac{v_{\rm DM}}{v} 
\right)^2
\frac{m_\nu}{M_1} = 1.3 \times 10^{-11}
\,.\label{Eq:mixing}
\end{eqnarray}
This relation is free from the scale of $v_{\FN}$ and the charge $n$. 
And, Eq. (\ref{Eq:mixing}) determines $v_{\rm DM}$ as 
\begin{eqnarray}
v_{\rm DM} \eqn{\simeq} 0.17\GeV\,.
\end{eqnarray}
This value can be obtained when $|m_3/m_2| \simeq 3.1 \times 10^{-2}$ 
in Eq. (\ref{Eq:vDM}).

\begin{figure}
\begin{center}
\includegraphics[width=6.5cm,clip,angle=-90]{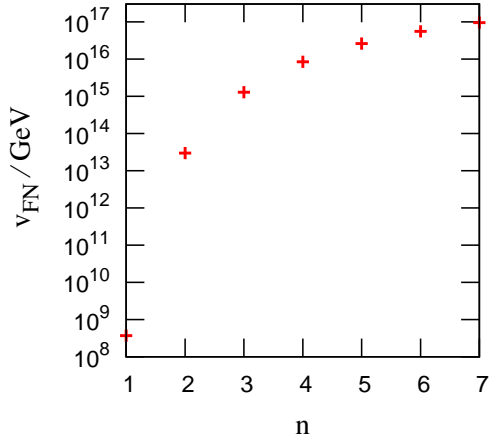}
\caption{The $n$ and $v_{\FN}$ plane. }
\label{FIG:n-vFN}
\end{center}
\end{figure}

Meanwhile, the remaining ambiguities of this model are the values of $v_{\FN}$ and $n$.
These values are also determined by the constraint of the DM mass.
The correlation between $v_{\FN}$ and $n$ is shown in Fig.~\ref{FIG:n-vFN}.
For example, when $n=1$, the breaking scale of $U(1)_{\FN}$ 
should be $\mathcal{O} (10^{8\mathchar`-9}) \GeV$ to explain DM properties.
On the other hand, when $n$ becomes large, the ordinary Froggatt-Nielsen 
mechanism \cite{Froggatt:1978nt} can work for generating the fermion mass hierarchy 
without any fine-tuning of the coupling constants.

We would like to give a brief comment on the production mechanism of DM.
One simple production mechanism is the so-called Dodelson-Widrow mechanism \cite{Dodelson:1993je} 
in which RH$\nu$DM is produced via the mixing after the Big Bang Nucleosynthesis.
However, the required magnitude of the mixing angle for this mechanism 
is much larger than the experimental bound. 
Thus, we need to consider other possibilities.
One possibility is the production via 
the B-L gauge boson s-channel exchange \cite{Khalil:2008kp}.
In this case, the relic abundance is evaluated as 
\begin{eqnarray}
\Omega_{\rm DM} h^2 \eqn{=} \Omega_{\nu_{R1}} h^2 = 
0.14 \left( \frac{M_1}{7 {\keV}} \right) \left( \frac{100}{g_\ast} \right)^\frac{3}{2} 
\left( \frac{T_R}{4.0 \times 10^{13} \GeV} \right)^3 \left( \frac{3.0 \times 10^{14} \GeV}{v_{\BL}} \right)^4 \,,\label{Eq:DM_abun}
\end{eqnarray}
where $h$ is the dimensionless Hubble parameter, 
and $T_R$ denotes the reheating temperature of the universe.
When we adopt the thermal production, the data 
from the Lyman-$\alpha$ forest \cite{Viel:2006kd}-\cite{Boyarsky:2008mt} should be taken 
into account.
However, this bound is 
relaxed by the late time entropy dilution~\cite{Bezrukov:2009th,Nemevsek:2012cd}.
The total amount of thermal abundance seems to be 
smaller than Eq.~(\ref{Eq:DM_abun}), 
but we can adjust the abundance by controlling the reheating temperature.
Although the sufficient amount of DM can also be explained in our setup, 
we skip the detailed estimation here.
Moreover, in this setup, we can also explain the baryon asymmetry of the universe via
the so-called leptogenesis scenario \cite{Raidal:2002xf} by non-thermal decay of the second lightest right-handed neutrino, $\nu_{R2}$. 
The lightest one is a candidate for DM as discussed above 
and the other right-handed neutrinos cannot decay into this lightest right-handed neutrino 
due to the difference of the $Z_2$ charge shown in Tab.~\ref{Tab:assign}. 
Such setup is the same as the split seesaw case \cite{Kusenko:2010ik}.

Lastly, we would like to add a comment on the detectability of RH$\nu$DM-philic Higgs boson, $H_{\rm DM}$.
$H_{\rm DM}$ pair production through the decay of $H$ is possible when the mass of 
$H_{\rm DM}$ is less than the half of the SM Higgs boson mass.
In this case, the invisible decay of the Higgs boson might be seen in the collider experiment 
depending on the strength of the portal couplings \cite{Haba:2011nb}.
On the other hand, $H_{\rm DM}$ is hardly produced because of its odd parity 
under $Z_2$ symmetry with $m_2 \gg m_1$. 
Detailed quantitative estimation is the future works at the moment.

\section{Summary}
We have suggested a new model in which 
the right-handed neutrino dark matter is naturally included.
It can explain the recently observed $3.5 {\keV}$ X-ray signal 
by the XMM-Newton observatory.
One of the most impressive features of our model is the existence of 
the dark matter-philic Higgs field, $H_{\rm DM}$.
The smallness of the left-right mixing angle of the lightest neutrinos 
is ensured by the smallness of the vacuum expectation value of this dark matter-philic Higgs field.
On the other hand, the smallness of the mass of the dark matter 
can be clearly understood by the Froggatt-Nielsen mechanism.
We have shown that the cosmological observations fix the scale of $v_{\rm DM}$ as $0.17\GeV$ 
and the intriguing point is that 
the uncertainty of this model to realize the DM mass is not essential to determine this scale.

%%%%%%%%%%%%%%%%%%%%%%%%%%%%%
\subsection*{Acknowledgement}
%%%%%%%%%%%%%%%%%%%%%%%%%%%%%

The authors thank to Alexander Merle and Kei Yagyu for giving us valuable comments.
This work is partially supported by Scientific Grant by Ministry of Education 
and Science, No. 24540272. The work of R.T. is supported by Research Fellowships
 of the Japan Society for the Promotion of Science for Young Scientists.

%%%%%%%%%%%%%%%%%%%%%%%%%%%%%%%%%%%%%%%%%%%%%%%%%%%%%%%%%%%%%%%%%%%%
%%%%%%%%%%%%%%%%%%%%%%%%%%%%%%%%%%%%%%%%%%%%%%%%%%%%%%%%%%%%%%%%%%%%
%%%%% ** Reference ** %%%%%%%%%%%%%%%%%%%%%%%%%%%%%%%%%%%%%%%%%%%%%%
%%%%%%%%%%%%%%%%%%%%%%%%%%%%%%%%%%%%%%%%%%%%%%%%%%%%%%%%%%%%%%%%%%%%
%%%%%%%%%%%%%%%%%%%%%%%%%%%%%%%%%%%%%%%%%%%%%%%%%%%%%%%%%%%%%%%%%%%%

%%%%%%%%%%%%%%%%%%%%%%%%%%%%%%%%%%%%%%%%%%%%%%%%%%%%%%%%%%%%%%%%%%%%
%%%%%%%%%%%%%%%%%%%%%%%%%%%%%%%%%%%%%%%%%%%%%%%%%%%%%%%%%%%%%%%%%%%%
%%%%%%%%%%%%%%%%%%%%%%%%%%%%%%%%%%%%%%%%%%%%%%%%%%%%%%%%%%%%%%%%%%%%
%%%%%%%%%%%%%%%%%%%%%%%%%%%%%%%%%%%%%%%%%%%%%%%%%%%%%%%%%%%%%%%%%%%%
\end{document}